\documentclass{article}

\usepackage{arxiv}

\usepackage[utf8]{inputenc} 
\usepackage[T1]{fontenc}    
\usepackage{hyperref}       
\usepackage{url}            
\usepackage{booktabs}       
\usepackage{amsfonts}       
\usepackage{nicefrac}       
\usepackage{microtype}      
\usepackage{lipsum}		
\usepackage{graphicx}
\usepackage{doi}
\usepackage{graphicx}
\usepackage{subcaption}
\usepackage{caption}
\usepackage{amsmath}
\usepackage{array}

\title{A NEW SCIENTIFIC INDEXING MODEL: U-INDEX}

\author{ \href{https://orcid.org/0000-0001-5359-1447}{\includegraphics[scale=0.06]{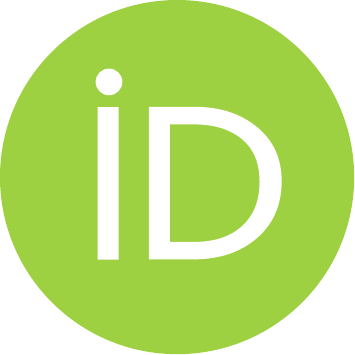}\hspace{1mm} Ugur Saglam}\thanks{Corresponding author}
    \\
	Department of Physics\\
	Istanbul University\\
	Istanbul, PA 34134 \\
	\texttt{usaglam@istanbul.edu.tr} \\
	\And
	\href{https://orcid.org/0000-0000-0000-0000}{\includegraphics[scale=0.06]{orcid.pdf}\hspace{1mm} Fatih Canata} \\
	Department of IR Management\\
	Istanbul University\\
	Istanbul, PA 34134 \\
	\texttt{fatihcan@istanbul.edu.tr} \\
}

\renewcommand{\shorttitle}{\textit{arXiv} Template}

\hypersetup{
pdftitle={A template for the arxiv style},
pdfsubject={q-bio.NC, q-bio.QM},
pdfauthor={David S.~Hippocampus, Elias D.~Striatum},
pdfkeywords={First keyword, Second keyword, More},
}

\begin{document}
\maketitle

\begin{abstract}
H-index has become more popular nowadays and is used for some scientific performance criteria in the world widely. This indexing method does not correctly measure any performance or carrier specifications because of the parameters that are used to form the measurement basis. H-index is located based on citation(C) and paper(N) parameters that involve no logical criterion on the counting process and so measurement on this basis can only give quantity results not any quality information. Therefore, we need a new indexing instrument to find out also the scientific quality unique to an individual author even if that takes into account the effect of multiple coauthorships. Ipso facto, we create a new bibliometric indicator or academic performance indicator called the u-index.
\end{abstract}

\keywords{Alternative indexing methods; Academic performance indicator; Impact factor; H-index; U-index.}

\section{Introduction}

Nowadays, it is getting important to quantify the scientific performance of authors. Research performance indicators mostly use bibliometric instruments based on parameters such as the number of papers, citations, or the impact factor of journals \cite{Leeuwen2003}. Papers are shared in scientific organizations such as journals, conferences, congresses, etc. with researchers and the public. Every scientific paper takes part in the scientific community by citing the former essays. The originality of papers is obtained by the number of citations and the scientific contribution. For the first time, Lotka \cite{Lotka1926} firstly suggested measuring the scientific performance by the number of papers, Gross and Gross \cite{Gross1927} suggested it as the number of citations. Garfield \cite{Garfield2006} founded Institute for Scientific Information, firstly mentioned the journal impact factor(JIF), classified citations into categories, and found out Science Citation Index listed the journals as to main motivations. JIF was planned to support the librarians in finding the impact of journals according to citations and the journals were started to array according to the impact factors in the advancing years \cite{Garfield1972}.

JIF is an assistant instrument to improve the core collection of librarians in earlier evolve to a different form as a dominant criterion in assignment and promotion, finding a job or a fund for projects, and measuring the performance of researchers \cite{Raan2005}. Specifically in the last years, considering bibliometric quantifying as the performance criterion of researchers has attracted attention to bibliometric instruments. The most popular metric instrument that is used to measure the individual performance of researchers and causes lots of discussions is h-index \cite{Zhang2012}. For instance, citation databases such as Web of Science, Scopus, and Google Scholar feature h-index instead of the other metrics \cite{Yong2014}. Hirsch \cite{Hirsch2005} published the essay on the h-index which is assumed to inform about the productivity and the impact of papers of a researcher. H-index can be defined as when an author has published h papers that have each been cited at least h times.

Hirsch \cite{Hirsch2005} states that the index is an important instrument to evaluate the researchers' competition for the same fund and sources. Since published the advantages of the h-index assumed as a dominant and common metric from different citation databases are as follows: H-index is the primary and intelligible combined metric to measure the individual performances of researchers; the metric combines the number of essay and citation impacts. H-index is a cumulative metric and there is no effect on the metric of increasing the number of papers; the h-index measures steady performances rather than fast growth; the metric that measures the scientific performance of a researcher may play an important role in academic promotions, research fund assignments, and scientific awards.

After the published h-index is criticized especially by the scientist that study bibliometrics. H-index due not to mind the publication year and the research area is accepted problematic to compare the people that work on different research areas and in different years \cite{Bornmann-Marx2014}. Though the dependency of publications and citations on the research lifetime, comparing the researchers from different academic lifetimes is considered a big mistake \cite{Bornmann-Marx2011,Glanzel2005,Glanzel2006}. Another problem of the h-index that suggests estimating the performance of an individual researcher is stated as that the metric does not notice the number of authors in a publication \cite{Hirsch2005}. Being not reasoned or nonsense to define the papers according to the citation impacts indicates that the h-index does not comply with the bibliometric standard \cite{Bornmann2014}. To lean h-index on not to be a logical relation between two parameters such as the number of papers and the number of citations is evaluated as an exceptional approximation according to bibliometric standard \cite{Bornmann-Marx2011}.

Nearly 37 different metric variants were suggested to overcome the problems and the deficiencies of h-index \cite{Bornmann-Mutz2011}. The variants evaluate the discipline difference, self-citations, co-authorship, career and publication lifetimes, and the cited publications except for the h-index core list also \cite{Bornmann-Mutz2011}. The fundamental aim of these models is to correct the deficiencies of the h-index and develop a comprehensive metric by changing the parameters such as research lifetime and co-authorship \cite{Alonso2009}.

\section{Method}

We have been in quest of a new bibliometric indicator or scientific performance index for the reasons mentioned in the previous section. We have worked on numerous bibliometric indexes but especially h-index and its variations that have the same methodological defects and arrived at some serious criticisms about the current performance indicators. The parameters of the measurement should be determined properly, even the parameters should be increased in number and the performance measuring methods should not affect the value of the performance indicator. Thus, an alternative scientific index has been developed over the basic parameters that can be easily obtained from the databases such as citation(C), the impact factor(IF), and paper count(N) for practical purposes. Moreover, a logical or so to say semantic condition has been put on the counting process of the measurement method.

The currently used method is called h-index using only citation parameters to align the papers. However, we think that the papers have to be aligned according to three parameters a citation(C), the impact factor(IF) of the journal, and the average citation/average impact factor ratio(CIF). In this way, we suppose that the indexing method can be more reliable and effective for a performance indicator. The alternative indexing method exhibits a critical behavior by considering the relation between citation and impact factor. In this way lots of tendencies of an author that wants to increase the number of citations can be explainable: by using the popularity factor of the journal, author, or field; preferring the most influential journals or collaborations; publishing a series of biased journal publications, attending the large scientific communities, being in the experimental science organizations. Such cases can cause an artificial positive bias on the h-index or the performance of an author. The academic performance indicator of an individual author has to be measured by some indexing instrument.

\begin{table}[h]
\begin{center}
\begin{tabular}{c|c|c|c|c|c}
\hline Author & $\mathrm{C}$ & $\mathrm{IF}$ & $\mathrm{C} / \mathrm{IF}$ & ${\mathrm{CIF}}$ & $\mathrm{N}(\mathrm{C} / \mathrm{IF} \ge {\mathrm{CIF}})$ \\
\hline pub1 & highest & ... & ... & const. & 1 \\
\hline pub2 & ... & ... & ... & const. & 2 \\
\hline ... &  &  &  &  & ... \\
\hline pubu & ... & ... & ... & const. & u \\
\hline ... &  &  &  &  & $-$ \\
\hline pubn & lowest & ... & ... & const. & $-$ \\
\hline
\end{tabular}
\caption{\label{tab:table-name}U-index}
\end{center}
\end{table}

The new model called as u-index finds out an academic performance indicator by using a new measure method dissimilar to the h-index. We can take on defining the h-index first, then the differences between the two models. The h-index is the maximum value of h such that the given author has published h papers and each paper has been cited at least h times. The index is designed to be developed via simpler measures such as the total number of citations or publications. Though citation routines differ widely among different fields, the index is supposed to work properly only for comparing scientists working in the same field. The u-index is the maximum value of u such that the given author has published u papers that have been ordered according to the citation count from highest to lowest one and each paper has the citation/impact factor ratio(C/IF) at least the average citation/average impact factor ratio(CIF) as in Table 1. This index is designed to be developed via simpler measures such as the total number of citations or publications and additionally the impact factor of the journals. Though citation routines differ widely among different fields, the index is supposed to work properly for comparing scientists working in all fields.

To find out the realistic personal performance indicator among scientist working in different fields, the natural publication algorithm have to be determined theoretically: the papers in the higher impact factor journals that exhibit higher average citation behavior must have higher citations and vice versa; the index must be the performance indicator for scientist working in different areas. The papers of an author should be ordered to correspond to the citation count for each publication as in h-index. Then, the last position in which the citation/impact factor of the journal ratio(C/IF) for each publication is greater than or equal to the average citation/average impact factor ratio(CIF) gives the index, as shown in Fig. 1. In this way, the index value relatively is getting closer and can be used for the performance indicator in all field.

\begin{figure}[h]
    \centering
    \includegraphics[width=0.60\textwidth]{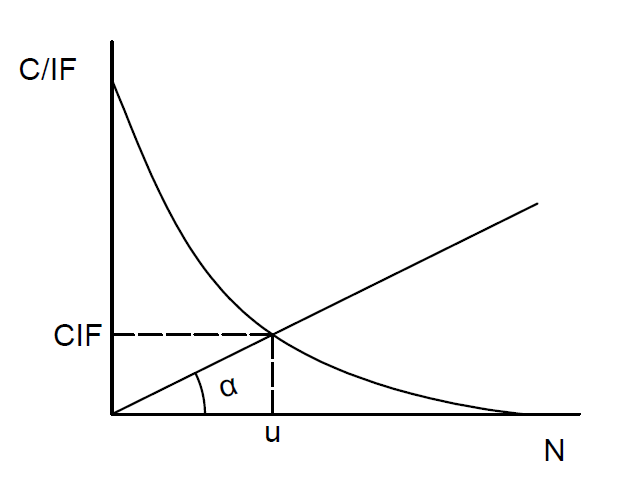}
    \caption{Citation/Impact Factor of Journal(C/IF)-Paper Number(N) Graph}
    \label{fig:my_label}
\end{figure}

U-index can be considered a powerful dynamical instrument to comment on the individual performances of researchers and fluctuates due to academic performance cyclically. The index tends to increase with new publications, decrease with citations to the current publications, and steady in the case of a balanced impact factor-citation ratio or no publication and citation. Thus this indicator exhibits a cyclical dynamical behavior according to the performance of researchers. Not being in the tendency to steadily increase is another important characteristic of the index and so we can mention a saturation behavior for every individual performance. The saturation of an index due to the end of the career of researchers is an essential characteristic of a performance indicator. This saturation point can be greater or lower than the index in the active years of a researcher.

We get the h-index from the u-index in the case of a good correlation between the citation count and the journal impact factors for each publication. The u-index can be considered a generalized form of the h-index. In fact, the h-index only measures good carrier development in agreement with the productivity and honesty of an author. Besides, the h-index is an excellent indicator of idealized carrier development behavior for individual researchers from all research areas. However, nowadays researchers exhibit enormous behavioral anomalies throughout their individual career lives to compete for academic promotions, research funds, and scientific awards. Thus u-index can determine the carrier anomalies and give chance to mend competition conditions.

The impact factor of journals is an effective parameter that determines the average citation of an essay published in a journal. Thus, an author's citation performance can be artificially affected by the impact factor or the average popularity of journals. When we want to take into account the pure scientific performance of an author, we have to distinguish the average citation factor of journals from the total number of citations. In this way, the vicious cycle that defines citation-impact factor correlation can be determined better. Citation-impact factor correlation can be defined as the number of citations that alter the journal impact factor or the number of average citations for a paper of an author and vice versa. The main point is that a paper should be published in a journal that minds the optimum scope and popularity conditions. In other words, the journal should not consider the popularity of authors in the refereeing and accepting process of papers.

Authors and also journals exhibit some behavioral anomalies to save and enhance their own index parameter and the current impact factor respectively. For authors: the desire to publish all the papers in the higher impact factor journals, as a result, want to attend the groups of senior authors, popular scientific communities, or large research groups. For journals: to accept lots of papers of senior authors and the collaborations or scholars of senior authors. The behavioral anomalies mentioned above rise as a consequence of accepting quantitative academic measurements as an indicator of academic performance.

The new model called u-index also foresees some futuristic outcomes. Every author minds the scope of journals and chooses the optimal one that corresponds to the scope, so the papers are published in the most relevant journals and so the popularity and the impact factor of a journal indicate better correlation. On the other hand, this model says if the scope and the impact factor are consistent, the impact factor and the citation count for a paper, so the citation count for a paper and the performance indicator index of an author will be consistent. Thus the impact factor of the journal in the same scope will be balanced and there will be no difference between the journal at the same scope.

U-index can be considered a new bibliometric instrument that aims to be an alternative point of view to the current problems of quantitative academic measurements. In some instances, there may be seen a huge difference in index value between researchers that work even in the same field. This is not a problem only related to the individual performances of researchers but also their strategic decisions to enhance their quantitative performance indicators. 

\section{Approach}

Metrics have a stress factor on authors for only productivity, not creativity or originality which can be seen on scientific performances mostly. Thus in the publication process journals and authors can exhibit several strategies to preserve their popularity through impact factor and h-index parameters. Thereupon it can be seen that some kinds of abnormalities related to authors and journals constitute the backbone of metric systems. We investigated the publication lists of numerous researchers, then presented some sample careers to obtain and discuss abnormal behaviors. We compile commonly seen author-level abnormal behaviors on some categories such as the effect of the senior author, large research group collaboration, senior co-author, and influential journal. 

In the case of the senior author effect, journals tend to accept papers from authors that are well-known in their own research areas. However, we notice a low C/high IF ratio when we check the publication lists of these authors. For the h-index, there is no problem as a performance indicator but the u-index determines this as an abnormality.

\begin{table}[h]
\begin{center}
\begin{tabular}{|m{2cm}|m{1cm}|m{1cm}|m{1.25cm}|m{0.75cm}|m{3cm}|}
\hline \textbf{Author a} & $\mathrm{C}$ & $\mathrm{IF}$ & $\mathrm{C} / \mathrm{IF}$ & ${\mathrm{CIF}}$ & $\mathrm{N}(\mathrm{C} / \mathrm{IF} \ge {\mathrm{CIF}})$ \\
\hline pub1 & 770 & 4,15 & 185,54 & 50 & 1 \\
\hline pub2 & 650 & 3,84 & 169,27 & 50 & 2 \\
\hline \textbf{pub3} & $\mathbf{120}$ & $\mathbf{6,15}$ & $\mathbf{19,51}$ & $\mathbf{50}$ & $\mathbf{-}$ \\
\hline pub4 & 100 & 1,86 & 53,76 & 50 & $-$ \\
\hline
\end{tabular}
\caption{\label{tab:table-name}Senior author effect}
\end{center}
\end{table}

Author a will reach a greater index value over u-index if publication 3 is not in the publication list of the author in Table 2. In this situation, the author uses its popularity to publish the paper in the journal with a higher impact factor, but not cited as expected. In fact, the essay should be published in a journal with less impact factor related to its popularity.

In the case of the large research group collaboration effect, junior authors tend to attend known research teams to raise their number of papers and citations. But we see an immediate decrease in the number of citations when we check the publication lists. For the h-index, all cited works should be assumed as performance indicators but the u-index determines abnormal publication-citation activities that are not coherent with the other papers in the publication list.

\begin{table}[h]
\begin{center}
\begin{tabular}{|m{2cm}|m{1cm}|m{1cm}|m{1.25cm}|m{0.75cm}|m{3cm}|}
\hline \textbf{Author b} & $\mathrm{C}$ & $\mathrm{IF}$ & $\mathrm{C} / \mathrm{IF}$ & ${\mathrm{CIF}}$ & $\mathrm{N}(\mathrm{C} / \mathrm{IF} \ge {\mathrm{CIF}})$ \\
\hline pub1 & 5200 & 3,56 & 1460,67 & 420 & 1 \\
\hline pub2 & 4160 & 4,88 & 852,46 & 420 & 2 \\
\hline pub3 & 3500 & 3,82 & 916,23 & 420 & 3 \\
\hline \textbf{pub4} & $\mathbf{1 8 0}$ & $\mathbf{2,12}$ & $\mathbf{8 4 , 9 1}$ & $\mathbf{4 2 0}$ & $\mathbf{-}$ \\
\hline pub5 & 100 & 1,88 & 53,19 & 420 & $-$ \\
\hline
\end{tabular}
\caption{\label{tab:table-name}Large research group collaboration effect}
\end{center}
\end{table}

Author b may reach a greater index value over the u-index if the first three publications are not in the publication list in Table 3. In this situation, author b uses the popularity of a scientific organization to have numerously cited papers. In fact, author b should have an index value through the papers only published with the research community that is not coherent with the other papers in the publication list.

In the case of the senior co-author effect, relatively junior authors tend to publish some papers with senior scientists in the related area to enhance their academic performances. But we see a critically low C/IF ratio in some papers even published in the same journal. For the h-index, there is only one parameter as C but u-index determines the abnormal correlation between C and IF in the publication list.

\begin{table}[h]
\begin{center}
\begin{tabular}{|m{2cm}|m{1cm}|m{1cm}|m{1.25cm}|m{0.75cm}|m{3cm}|}
\hline \textbf{Author c} & $\mathrm{C}$ & $\mathrm{IF}$ & $\mathrm{C} / \mathrm{IF}$ & ${\mathrm{CIF}}$ & $\mathrm{N}(\mathrm{C} / \mathrm{IF} \ge {\mathrm{CIF}})$ \\
\hline pub1 & 120 & 4,58 & 26,20 & 12 & 1 \\
\hline pub2 & 100 & 4,16 & 24,04 & 12 & 2 \\
\hline \textbf{pub3} & $\mathbf{20}$ & $\mathbf{4,16}$ & $\mathbf{4,81}$ & $\mathbf{12}$ & $\mathbf{-}$ \\
\hline pub4 & 15 & 4,58 & 3,28 & 12 & $-$ \\
\hline
\end{tabular}
\caption{\label{tab:table-name}Senior co-author effect}
\end{center}
\end{table}

Author c wants to reach a greater index value via the first two publications over the u-index as seen in the publication list in Table 4. In this situation, author c uses the popularity of a co-author to boost their career via the journal with a high impact factor. In fact, author c uses the opportunity factor to develop good scientific relations and give accelerate career development in the later years of the research lifetime.

In the case of the influential journal effect, authors tend to publish their research in the journal with the highest impact factor to increase their index value. But we see  commonly a low C/high IF ratio in this type of journal/paper correlation. For the h-index, there is no problem as long as the number of citations is higher but the u-index notices the journals with higher impact factors.

\begin{table}[h]
\begin{center}
\begin{tabular}{|m{2cm}|m{1cm}|m{1cm}|m{1.25cm}|m{0.75cm}|m{3cm}|}
\hline \textbf{Author d} & $\mathrm{C}$ & $\mathrm{IF}$ & $\mathrm{C} / \mathrm{IF}$ & ${\mathrm{CIF}}$ & $\mathrm{N}(\mathrm{C} / \mathrm{IF} \ge {\mathrm{CIF}})$ \\
\hline pub1 & 2400 & 3,56 & 674,16 & 152 & 1 \\
\hline pub2 & 2200 & 4,18 & 526,32 & 152 & 2 \\
\hline pub3 & 1400 & 2,88 & 486,11 & 152 & 3 \\
\hline pub4 & 1000 & 3,12 & 320,51 & 152 & 4 \\
\hline \textbf{pub5} & $\mathbf{850}$ & $\mathbf{39,22}$ & $\mathbf{21,67}$ & $\mathbf{152}$ & $\mathbf{-}$ \\
\hline pub6 & 780 & 3,48 & 224,14 & 152 & $-$ \\
\hline
\end{tabular}
\caption{\label{tab:table-name}Influential journal effect}
\end{center}
\end{table}

Author d wants to publish some essays in journals with a higher impact factor to reach a greater index value via the h-index as seen in the publication list in Table 5. In this situation, author d uses the acceleration of their career to publish in the most influential journals. In fact, author d takes advantage of its papers with numerous citations to win admittance from prestigious journals.

Here we sum up the author-level abnormal behaviors, and mention sometimes potential journal-level abnormal behaviors that are mostly an intricate correlation between authors and journals. H-index may be used as a tool by most authors for the purpose of academic promotions, research fund assignments, and scientific awards. Therefore a new metric that is not abuse-liable has to be developed as a more reliable academic performance indicator. U-index exhibits a behavior-responsive algorithm that may be expressed artily as having semantic awareness in other words.

\section{Conclusion}

U-index can be considered as a count criterion on the h-index that restricts the numbering according to the average value(CIF). Measuring academic performance via a bibliometric indicator may be seen as impossible or not a quantitative problem as we consider nowadays. However, researchers need to be assessed via positive measurement systems such as bibliometric instruments that want to be used for academic promotions, research fund assignments, and scientific awards. Though the quantity-quality dilemma is seen as not to be solved by any bibliometric instruments, we want to determine an academic performance indicator over the basic parameters of the measurement such as citation(C), the impact factor(IF), and average citation/average impact factor ratio(CIF) that can be obtainable from the platforms providing scientific and academic data, information and analytics. 
U-index is a semi-semantic algorithm that determines the abnormal behaviors of researchers and helps to identify them. Four common types of behavioral abnormality have been detected in the publication lists among numerous researchers and the situations are ensampled separately to identify the motivation of each type. The common angst as known is to get a good academic career opportunity, high index value and so some kind of promotion, fund, and awards. Thus we can surmise that nearly all authors from all research fields feel the same pressure of getting high index values in the process of their career development. This is becoming a more common type of anxiety among early career researchers and even senior ones that do not have a relatively satisfying career among contemporaries in evidence.

Measuring academic performance should be a positive stimulant and an analysis not only quantitative but also qualitative. H-index is cited mostly as an example because of its popularity as a pure quantitative analyzer and the source of aforementioned anxiety of researchers also. The prior and present metric models that are created in the basic parameters and some mathematical instruments have not responded to the need for performance measurement standards. At this point, the need to develop a proper indicator of academic performance is the main motivation of this paper and so we create a model called u-index that constitutes simply accessible parameters from some databases. We have analyzed the u-index by carefully working on goodly numerous researchers, obtained some author-level abnormality categories, monitored as tables, and discussed in detail therein before.






\bibliographystyle{unsrt}
\bibliography{references}  






\end{document}